# Dominant shear bands observed in amorphous ZrCuAl nanowires under simulated compression

**Qiran Xiao**, Department of Materials Science and Engineering, Rensselaer Polytechnic Institute, Troy, New York 12180
**H.W. Sheng**, School of Physics, Astronomy and Computational Sciences, George Mason University, Fairfax, Virginia 22030
**Yunfeng Shi**, Department of Materials Science and Engineering, Rensselaer Polytechnic Institute, Troy, New York 12180
Address all correspondence to Yunfeng Shi at **shiy2@rpi.edu**

## Abstract

We observed the formation of dominant shear bands in model ZrCuAl metallic glass (MG) nanowires (18-nm-long) in molecular dynamics simulations, which implies size-independent incipient plasticity in MG materials. The MG nanowires were prepared using the simulated casting technique to ensure proper relaxation of sample surfaces. Under uniaxial compression, shear bands initiate at the surfaces and lead to reduced icosahedral short-range order. The shear band formation is sensitive to sample thermal-history, which calls for careful consideration of sample preparation effects in both experimental and numerical studies of size-effect in MG samples.

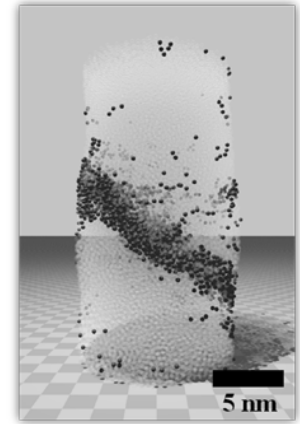

5 nm

Metallic glasses (MG), unlike other forms of glasses, have distinctive mechanical properties including large elastic strain limit and high tensile strength [1,2]. Metallic glasses, however, suffer from catastrophic failure upon free propagating shear bands. The formation of localized shear bands strongly compromises the practical usages of the metallic glasses for load-bearing applications. One promising way to overcome this shortcoming of metallic glasses is to reduce the sample size of MGs using Focused Ion Beam (FIB) [3] or hot extension [4]. It has been shown that MG in the form of nanowires can exhibit large tensile strain [5,6]. This is thought to be due to the limited spatial scale of nanowires which is insufficient for shear band propagation and maturation into run-away cracks [1].

Although shear band propagation and maturation into cracks are likely size-dependent [1], it is still not clear whether the initiation of shear bands (incipient plasticity) is size-dependent. Both size-dependent [3,7-10] and size-independent [12-14] shear banding behaviors in MG nanorods have been reported in experiments. Moreover, MD simulations on two-dimensional or three-dimensional thin-slab geometry [15,16] show clear shear band formation, while MD simulations on three-dimensional nanowire samples show homogeneous flow [17,18]. Recently, Shi [11] proposed a new numerical sample preparation procedure termed "simulated casting", during which a molten liquid confined in a cylindrical container is quenched directly into a glassy nanowire. This sample preparation technique is advantageous in comparison to the conventional cut-form-bulk procedure to obtain nanowire samples in computer simulation. Surfaces resulting from the traditional cut-from-bulk procedure are usually not sufficiently relaxed at the processing temperature, whereas in the simulated casting technique, surfaces are created from the melt directly and atoms on the surfaces are fully relaxed. For MG nanowires obtained from the cut-from-bulk procedure, homogeneous deformation followed by necking was observed owing to unrelaxed surfaces [11]. By contrast, it was shown that MG nanowires obtained from the simulated casting, thus with relaxed surfaces, exhibit shear banding with a sample length longer than 13 nm, as a result of competition between elastic energy and shear band energy [11]. Interestingly, this critical length value is very close to experimental reports of critical shear band nucleus length of 11 nm [8] and a critical defect length of 15 to 25 nm [19].

It is useful to note that a Wahnstrom binary Lennard-Jones (LJ) system was used in our prior study. Thus, an interesting question to ask is whether the observed size-independent shear banding arises from the simplicity of the interatomic force field. In this work, a realistic ZrCuAl tertiary system with many-body embedded atom method (EAM) potential was selected as the simulation system. ZrCuAl nanowires were virtually synthesized using the simulated casting technique [11] and then subjected to nanomechanical tests followed by structural and deformation analysis. Uniaxial compression, instead of tensile tests previously used, was conducted to better reflect the majority of experimental testing conditions on MG nanopillars (compression under a flat punch). Simulations on ZrCuAl nanowires with a length of 17 nm exhibit dominant shear bands during compression, which agrees with previous simulations using LJ systems. It was found that shear bands initiate at the surface and destroy icosahedral clusters while propagating. Similar to bulk samples, shear bands can be suppressed by increasing the cooling rate for the vitrification of MG. Thus, the thermal history of MG nanowires also affects shear band formation, which should be considered in evaluating whether the localized deformation mode in MG is size-dependent.

We employed a newly-developed EAM potential for ZrCuAl ternary system [20]. The force field parameters were fitted to match *ab initio* calculations of over thousands of configurations including various crystalline, liquid and glassy states. This ZrCuAl EAM potential has been validated against a large array of experimental and *ab inito* data, and has been extensively used to investigate the atomic-level structure, formation and deformation mechanisms of ZrCuAl bulk metallic glasses [16,18,20].

ZrCuAl nanowires were prepared using the simulated casting procedure as follows. The initial system consists of 49320 atoms with a Zr:Cu:Al atomic number ratio of 5:4:1. The ZrCuAl liquid was equilibrated at 2000K and was confined within a cylindrical wall with a density of 6.76 g/cm$^3$. The wall interacts repulsively with all atoms regardless of species. The repulsive potential follows $k/2(r-r_0)^2$, where $r$ is the distance between the atom and the cylindrical axis, while $r_0$ is the radius of the wall and $k$ is the spring constant. The interaction force between the atom and the wall is zero if $r$ is less than $r_0$. In this work, $r_0$ was set to be about 4.0 nm, and the initial spring constant k was set to 160.2 kg/s$^2$. This spring constant decreases to zero at the end of cooling to achieve zero radial-stress for the nanowires.

The ZrCuAl liquid confined by the repulsive wall was then cooled from 2000 K to 300 K in the canonical ensemble, which is analogous to the experimental casting procedure. Nanowires from such a "simulated casting" procedure is advantageous over those obtained from the traditional cut-from-bulk method to sculpturing MG nanowires [17,18]. Surfaces of the nanowires are allowed to relax as much as the interior of the nanowires. Three different quenching rates were used: 17, 170 and 1700 K/ns, respectively. The lowest quenching rate is very close to the one used in Ref. [20]. For each quenching rate, three independent samples were generated to examine sample-to-sample variations. All samples are subjected to periodic boundary conditions (PBC) along the cylindrical axis direction. Thus-formed MG nanowire samples are about 7.8 nm in diameter and 17.7 nm in length.

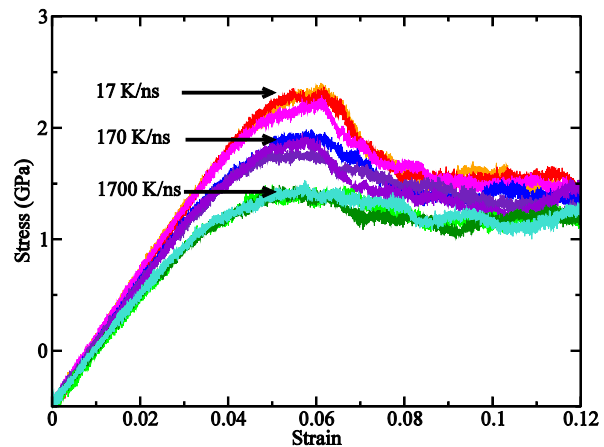

**Figure 1**. Stress-strain curves for ZrCuAl nanowires with three quenching rates under uniaxial compression tests. There are three independent samples for each quenching rate which are clustered together.

Uniaxial compression tests were conducted without temperature control on the ZrCuAl MG nanowire samples at a constant strain rate of 0.0001 ps$^{-1}$ to 12% nominal strain. Figure 1 shows the stress-strain curves for compression simulations on nanowires with three distinct quenching rates each with three independent samples. Note that considerable tensile stress was developed at the end of cooling along the nanowire axis. Such residual stress is not unexpected considering the mismatch of thermal expansion coefficients between the metallic liquid and the container (exactly zero in this simulation setup). The averaged Young's modulus for each quenching rate is 60 GPa, 55 GPa and 49 GPa respectively, in the order of increasing quenching rate. Note that the experimental Young's modulus for $Zr_{50}Cu_{40}Al_{10}$ is 88 GPa [22]. Since the quenching rates in MD simulations are much higher than experiments, it is not surprising that the Young's modulus is lower than the experimental value. In addition, the presence of surfaces may also affect Young's modulus significantly in such small amorphous nanowires. The ultimate yield stresses are 2.3 GPa, 1.9 GPa and 1.5 GPa respectively, from slow to fast quenching. The nanowires made with a slower quenching rate show a pronounced yield behavior which corresponds to the formation of shear bands. On the other hand, nanowires with higher quenching rates show almost elastic-perfect-plastic behavior indicating homogeneous plastic flow. These observed elastic and plastic deformation behaviors for amorphous nanowires are almost identical to those observed for bulk amorphous materials, that is, they are highly sensitive to the quenching rate for sample preparation. To further confirm that the difference in the plastic deformation behavior is due to the quenching rate, the local plastic strain for all amorphous nanowire samples was calculated at 12% nominal strain (Figure 2). Nanowires with low quenching rates exhibit dominant shear bands, as shown both on the surfaces and on the cross-sections, while nanowires with high quenching rates exhibit mostly homogeneous deformation without localization. One difficulty for both computational and experimental observation of the plastic event is to identify atomic activities hidden beneath the surface for three-dimensional systems. Figure S1 shows a transparent view of one MG nanowire with low quenching rate, which demonstrates that there is no hidden plastic activity outside the main shear band.

Next, we investigated whether the shear band initiates on the surfaces or inside the bulk, which is crucial to understand the surface-effect on shear band nucleation. For nanowires with a quenching rate of 17 K/ns, Figure S2 shows all atoms that participate in the plastic deformations of the nanowires along the axis direction, while hiding the undeformed atoms (atoms with a local deviatoric strain lower than 20%). The range of nominal compressive strain is from 6% to 8%, which roughly corresponds to the stress drop in the stress-strain curves as in Figure 1. For all the three cases, deformation initiates at the surfaces and then expands to the entire cross-section of the nanowire. Thus, the nucleation of shear bands is very sensitive to the surfaces of nanowires. This is in agreement with the observation that unrelaxed surfaces lead to multiple shear band nucleation sites and thus suppress a dominant shear band [11].

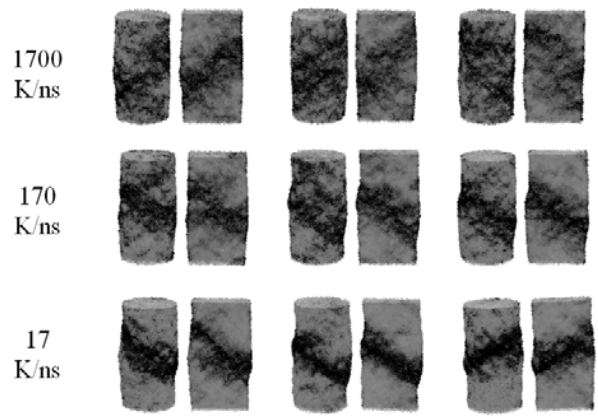

**Figure 2**. Deformation morphologies for ZrCuAl nanowires with three quenching rates under uniaxial compression tests at 12% nominal strains. Gray indicates zero local shear strain while black indicates 35% local shear strain.

Lastly, we examined the short-range order (SRO) of the material inside and outside the shear band [15]. Figure S3 shows the projected population of icosahedral centers as a distance perpendicular to the shear band position. The projection direction is selected perpendicular to the shear band at 12% nominal strain. It is clear that the icosahedral center population is uniform at 0% nominal strain and drops significantly at 12%. The half-height width of the shear band at 12% strain is about 3 nm. As the icosahedral SRO is considered the backbone of MG [21], the depletion of such local clusters leads to softening. Together with the observation that there are no thermal spikes associated with the shear band, the formation of the shear band is due to structural softening instead of thermal softening. The above observations in three-dimensional MG nanowires are very similar to molecular dynamics simulation results on MG thin-slab samples in two-and-half dimensions [15].

It is important to recognize the limitations of MD simulation in terms of quenching rate and strain rate, which are orders of magnitude higher than typical experimental values. One effective simulation strategy to mitigate the limitations is to investigate the strain-rate/quench-rate dependency by carrying out simulations at different strain-rates/quench-rates. From Figure 2, it is apparent that the samples with a high quenching rate deform homogeneously, while the samples with a low quenching rate exhibit shear banding. If one extrapolates such quenching rate dependency to the experimental condition (effectively zero quenching rate in simulations), dominant shear bands will form in MG nanowires. A similar trend has been observed computationally, and experimentally (embrittlement due to low cooling-rates [23] and embrittlement due to annealing [24]). Regarding the strain-rate effect, Figure S4 shows the deformation morphologies of uniaxial compression tests with a low strain rate of $0.00001 \ ps^{-1}$, which are very similar to Figure 2. It has been shown experimentally [25] that lower strain rates lead to larger shear band spacings (i.e., more localized deformation). Thus, shear banding is expected to be the main deformation mode for MG nanowires with very low quenching rate under very low strain rates.

In summary, model ZrCuAl MG nanowires prepared from the simulated casting technique show shear banding at a length of 17.7 nm and a radius of 3.9 nm. As the radius of the MG nanowire is comparable to the shear band width ~3 nm, our results support size-independent incipient plastic deformation mode for MG materials. The glassy state of experimental MG nanowires will be affected structurally and chemically during sample preparation, for instance, focused ion beam irradiations. The glassy state of the nanowires of small diameters is likely to be affected more than those with large diameters due to larger surface to volume ratio, which may help render apparent size-dependency from sample preparation. Careful consideration of sample preparation effects may help resolve the inconsistency in experimental observations on size-dependency in MG systems.

## Acknowledgments

We thank Michael Falk, Craig Maloney, Liping Huang, Despina Louca and Peter K. Liaw for stimulating discussions. YFS thanks the financial support from NSF under Grant No. CMMI- 1031408. HWS acknowledges financial support from NSF under Grant No. DMR-0907325. The molecular dynamics simulations are carried out in LAMMPS using computational facilities in the Computational Center for Nanotechnology Innovations (CCNI) at RPI.

## Supplementary materials

For supplementary material for this article, please visit http://dx.doi.org/10.1557/mrc.2011.26

## References

[1]  C. Schuh, T. Hufnagel, and U. Ramamurty: Mechanical behavior of amorphous alloys. *Acta Materialia* **55**, 4067-4109 (2007).

[2]  A.L. Greer: Metallic glasses...on the threshold. *Materials Today* **12**, 14-22 (2009).

[3]  C.A. Volkert, A. Donohue, and F. Spaepen: Effect of sample size on deformation in amorphous metals. *J. Appl. Phys.* **103**, 083539 (2008).

[4]  K. Nakayama, Y. Yokoyama, T. Ono, M. Chen, K. Akiyama, T. Sakurai, and A. Inoue: Controlled Formation and Mechanical Characterization of Metallic Glassy Nanowires. *ADVANCED MATERIALS* **22**, 872 (2010).

[5]  D. Jang, and J.R. Greer: Transition from a strong-yet-brittle to a stronger-and-ductile state by size reduction of metallic glasses. *Nat Mater* **9**, 215-219 (2010).

[6]    H. Guo, P.F. Yan, Y.B. Wang, J. Tan, Z.F. Zhang, M.L. Sui, and E. Ma: Tensile ductility and necking of metallic glass. *Nat Mater* **6**, 735-739 (2007).

[7]    A.R. Yavari, K. Georgarakis, W.J. Botta, A. Inoue, and G. Vaughan: Homogenization of plastic deformation in metallic glass foils less than one micrometer thick. *Phys. Rev. B* **82**, 172202 (2010).

[8]    A. Bharathula, S.W. Lee, W.J. Wright, and K.M. Flores: Compression testing of metallic glass at small length scales: Effects on deformation mode and stability. *Acta Materialia* **58**, 5789-5796 (2010).

[9]    D. Jang, C.T. Gross, and J.R. Greer: Effects of size on the strength and deformation mechanism in Zr-based metallic glasses. *International Journal of Plasticity* **27**, 858-867 (2011).

[10]   C.Q. Chen, Y.T. Pei, O. Kuzmin, Z.F. Zhang, E. Ma and J. Hosson: Intrinsic size effects in the mechanical response of taper-free nanopillars of metallic glass. *Phys. Rev. B* **83**, 180201 (2011).

[11]   Y. Shi: Size-independent shear band formation in amorphous nanowires made from simulated casting. *Appl. Phys. Lett.* **96**, 121909 (2010).

[12]   X.L. Wu, Y.Z. Guo, Q. Wei, and W.H. Wang: Prevalence of shear banding in compression of Zr41Ti14Cu12.5Ni10Be22.5 pillars as small as 150 nm in diameter. *Acta Materialia* **57**, 3562-3571 (2009).

[13]   A. Dubach, R. Raghavan, J.F. Löffler, J. Michler, and U. Ramamurty: Micropillar compression studies on a bulk metallic glass in different structural states. *Scripta Materialia* **60**, 567-570 (2009).

[14]   B.E. Schuster, Q. Wei, T.C. Hufnagel, K.T. Ramesh: Size-independent strength and deformation mode in compression of a Pd-based metallic glass. Acta Mater. 56, 5091-5100 (2008).

[15]   Y. Shi, and M. Falk: Atomic-scale simulations of strain localization in three-dimensional model amorphous solids. *Phys. Rev. B* **73**, (2006).

[16]   Y.Q. Cheng, A.J. Cao, and E. Ma: Correlation between the elastic modulus and the intrinsic plastic behavior of metallic glasses: The roles of atomic configuration and alloy composition. *Acta Materialia* **57**, 3253-3267 (2009).

[17]   Q.K. Li, and M. Li: Molecular dynamics simulation of intrinsic and extrinsic mechanical properties of amorphous metals. *Intermetallics* **14**, 1005-1010 (2006).

[18]   Y.Q. Cheng, A.J. Cao, H.W. Sheng, and E. Ma: Local order influences initiation of plastic flow in metallic glass: Effects of alloy composition and sample cooling history. *Acta Materialia* **56**, 5263-5275 (2008).

[19]   C.J. Lee, Y.H. Lai, J.C. Huang, X.H. Du, L. Wang, and T.G. Nieh: Strength variation and cast defect distribution in metallic glasses. *Scripta Materialia* **63**, 105-108 (2010).

[20]   Y. Cheng, E. Ma, and H. Sheng: Atomic Level Structure in Multicomponent Bulk Metallic Glass. *Phys. Rev. Lett.* **102**, (2009).

[21]   Y. Shi, and M. Falk: Does metallic glass have a backbone? The role of percolating short range order in strength and failure. *Scripta Materialia* **54**, 381-386 (2006).

[22]   Y. Yokoyama, T. Yamasaki, P.K. Liaw, and A. Inoue: Study of the structural relaxation-induced embrittlement of hypoeutectic Zr–Cu–Al ternary bulk glassy alloys. *Acta Materialia* **56**, 6097-6108 (2008).

[23]   X.J. Gu, S.J. Poon, G.J. Shiflet, and J.J. Lewandowski: Ductile-to-brittle transition in a Ti-based bulk metallic glass. *Scripta Materialia* **60**, 1027-1030 (2009).

[24]   G. Kumar, D. Rector, R.D. Conner, and J. Schroers: Embrittlement of Zr-based bulk metallic glasses. *Acta Materialia* **57**, 3572-3583 (2009).

[25]   H. Zhang, S. Maiti, and G. Subhash: Evolution of shear bands in bulk metallic glasses under dynamic loading. *Journal of the Mechanics and Physics of Solids* **56**, 2171-2187 (2008).